\documentclass[pra,twocolumn,groupedaddress,showpacs]{revtex4}
\usepackage{graphics}
\usepackage{enumerate}
\begin{document}
\newcommand{\beq}{\begin{equation}}
\newcommand{\eeq}{\end{equation}}
\newcommand{\beqa}{\begin{eqnarray}}
\newcommand{\eeqa}{\end{eqnarray}}
\bibliographystyle{apsrev}

\title{On repeated(continuous) weak measurements of a single copy of an unknown quantum state.}

\author{N.D. Hari Dass}
\email{dass@tifrh.res.in }
\affiliation{TIFR-TCIS, Hyderabad 500075, INDIA.}


\begin{abstract}
In this paper we investigate repeated
weak measurements,without post-selection, 
on a \emph{single copy} of an \emph{unknown} quantum state. 
The resulting random walk in state space is precisely characterised in terms of joint probabilities for outcomes.
We conclusively answer, in the negative, the very important question whether the statistics of such repeated measurements
can determine the unknown state. We quantify the notion of error in this context as the departure of a suitably averaged density matrix 
from the initial state. When the number of weak measurements is small the original state is preserved to a great degree, but only an
ensemble of such measurements, of a complete set of observables, can determine the unknown state.
By a careful analysis of errors, it is shown that there is a precise tradeoff between errors and \emph{invasiveness}. Lower the errors, greater
the invasiveness. Though the outcomes are not independently distributed,
an analytical expression is obtained for how averages are distributed, which is shown to be the way outcomes are distributed in a \emph{strong
measurement}. An \emph{error-disturbance} relation, though not of the Ozawa-type, is also 
derived. In the limit of vanishing errors, the invasiveness approaches what would obtain from strong measurements.

\end{abstract}

\maketitle

\section{Introduction}

Much of the original thinking about measurements in Quantum Theory is intimately tied up with the measurements being of the 
original von Neumann-Dirac type, also
known as \emph{strong} or more precisely, \emph{Projective Measurements}. Since then, many new and novel types of measurement schemes
have got unveiled like \emph{Protective Measurements}, \emph{POVM Measurements}, \emph{Weak Measurements} etc \cite{wisemannbook2010}. The foundational
aspects have to be revisited within the context of each of these measurement schemes. In this paper, we do so for Weak Measurements 
without \emph{Post-Selection}, and in particular address the issue of repeated such measurements on a \emph{single} copy. We shall use
the von Neumann-Dirac formalism to describe these measurements \cite{AVA;PRL1988,AV;PRA1989}.
\section{Weak Measurements Without Post-Selection}
Let S be the observable of the system with $s_i,|s_i\rangle_S$ its spectrum, which we take to be
\emph{non-degenerate}. 
The initial states of the system and the apparatus are taken to be \emph{pure}. Generalization to 
the \emph{mixed} case should be straightforward. For ensemble measurements, such generalized treatments are already available
in \cite{kofman2012}. 

Let $|\psi\rangle_S=\sum_i\,\alpha_i\,|s_i\rangle_S$ be the \emph{unknown} initial state of the system on 
which measurements of S are done. The expectation value of S in the state $|\psi\rangle_S$ is given by $\langle\psi|S|\psi\rangle_S=
\sum_i\,|\alpha_i|^2\,s_i$. The \emph{Pointer States} of the apparatus denoted by $|p\rangle_A$, are taken to be eigenstates of
an apparatus observable $P_A$. The point of view taken here is that such pointer states form the basis in which the density matrix 
becomes diagonal as a result 
of \emph{decoherence}. They are not always labelled by the mean values of $P_A$ in a given state of the apparatus.
Therefore, the specification of an \emph{apparatus} involves some quantum system, along with a decoherence mechanism which picks out
the pointer states.

In the original models the \emph{initial} apparatus state had to be necessarily
a pointer state and the system-apparatus interaction had to be of the type $H_I= g(t)\lambda\,S\,Q_A$ where $Q_A$ was \emph{canonically conjugate}
to $P_A$ i.e $[Q_A,P_A]=i\hbar$. In \cite{kofman2012} one can find extensive discussion of how to go beyond such restrictions as far as
weak measurements are concerned. But we shall stick to this choice.
More pragmatically, in the original model the initial aparatus states were
taken to be \emph{sharp Gaussian states} centred around some $p_0$. In other words, for $p_0=0$,
\begin{equation}
|\phi_0\rangle_A = N\,\int\,dp\,e^{-\frac{p^2}{2\Delta_p^2}}\,|p\rangle_A\quad\quad N^2\sqrt{\pi\Delta_p^2}=1
\label{eq:iniapp}
\end{equation} 
In the case of projective measurements, $\Delta_p << 1$. For weak measurements, however, $\Delta_p >> 1$.That means that the initial apparatus state is a \emph{very broad} superposition of pointer states with practically \emph{equal} weight for
each pointer state. We shall see that a combination of these factors results, with high probability, in neither the apparatus nor the system 
changing appreciably as a result of the measurement. Hence the name \emph{Weak Measurements}.

In both strong and weak measurements, the measurement interaction is taken to be \emph{impulsive} i.e the function g(t) is
nonvanishing only during a very small duration, say, $-\epsilon < t < \epsilon$. Without loss of generality g(t) can be taken to satisfy 
$\int\,dt\,g(t)=1$, and $\lambda=1$. The impulsive approximation is clearly an idealisation not shared by real life measurements. It is easy to work 
out the combined (pure)state of the system and apparatus after the impulsive measurement interaction is complete:
\begin{equation}
|\Psi(t>\epsilon)\rangle_{SA} = N\int\,dp\,\sum_i\,\alpha_i\,e^{-\frac{(p- s_i)^2}{2\Delta_p^2}}\,|s_i\rangle_S\,|p\rangle_A
\label{postMI}
\end{equation}
When $\Delta_p << 1$, only $p\approx\,s_i$ dominate and one recovers the well known results for strong measurements. As has been
well emphasized, this is still an \emph{entangled} state of the system and apparatus, and therefore does not reflect the fact that 
measurements have definite outcomes i.e after the measurement has been completed, the apparatus should be left in only one of the pointer
states. It is believed that \emph{environmental decoherence} diagonalizes the resulting combined pure density matrix
in the pointer states basis, 
to give
the \emph{post-measurement} density matrix: 
\begin{equation}
\rho^{post}_{SA} = \int dp\, |N(p,\{\alpha\})|^2|p\rangle\langle p|_A\,|\psi(p,\{\alpha\}\rangle\langle \psi(p,\{\alpha\}|_S
\label{eq:weakpost}
\end{equation}
where
\begin{equation}
N(p,\{\alpha\}) = N\,\sqrt{\sum_i\,|\alpha_i|^2\,e^{-\frac{(p- s_i)^2}{\Delta_p^2}}}
\label{eq:Nweakpost}
\end{equation}
\begin{equation}
|\psi(p,\{\alpha\}) = \frac{N}{N(p,\{\alpha\})}\,\sum_j\,\alpha_j\,e^{-\frac{(p- s_j)^2}{2\Delta_p^2}}\,|s_j\rangle_S
\label{eq:psiweakpost}
\end{equation}
Several aspects of weak measurements, which have puzzled many(for a very detailed account of various aspects of such measurements, see 
\cite{kofman2012}) , can be clarified with the help of eqns.(\ref{eq:weakpost},\ref{eq:Nweakpost},
\ref{eq:psiweakpost}). The measurement process gets completed when a single output of the apparatus, say, p, and a single system state 
$\psi(p,\{\alpha\})\rangle_S$ are picked from the mixture of eqn.(\ref{eq:weakpost}). The outcome p can, first of all, range over $[-\infty,\infty]$, far beyond the eigenvalue-range of S. 
Since the associated system state is in general not an eigenstate of S, there is no meaning to associating any 'value' of the observable to p.
System states corresponding to different outcomes are not \emph{orthogonal}, yet the measurements are of the POVM type, with the measurement
operators $M_p$ 
given by
\begin{equation}
\label{eq:weakpovm}
M_p = N\,\sum\limits_i\,e^{-\frac{(p-s_i)^2}{2\Delta_p^2}}
\end{equation}

For an \emph{ensemble} of weak measurements, $P(p,\{\alpha\})=|N(p,\{\alpha\})|^2$ being the probability for outcome p, the mean outcome is
\begin{equation}
\langle p \rangle_\psi = \int dp\, p|N(p,\{\alpha\})|^2 = \sum_i\,|\alpha_i|^2\,s_i 
\label{eq:weakmean}
\end{equation}
yielding the same expectation value as in strong measurements. Therefore \emph{state tomography}(such tomography using
weak measurements with \emph{post-selection} are considered in \cite{lundeen2011,swu2013};
repeated weak measurements as a means of augmenting projective tomography is considered in \cite{arvind2014}.) can be achieved through such ensemble weak
measurements. The variance of the outcomes can be readily calculated to yield
\begin{equation}
(\Delta p)^2_\psi = (\Delta p)^2+(\Delta S)^2
\label{eq:weakvariance}
\end{equation}
This exposes one of the major weaknesses(!) of weak measurements i.e the errors in individual measurements are huge. This can be 
reduced statistically as usual. If one considers averages over $M_w$ measurements, the variance in the average, 
is $\frac{\Delta_p}{\sqrt{2M_w}}$. It makes sense to compare different measurement schemes only for a \emph{fixed} statistical error.
Therefore if averaging is done over $M_s$ strong measurements, 
\begin{equation}
\frac{\Delta S}{\sqrt{M_s}}=\frac{\Delta_p}{\sqrt{2M_w}}\rightarrow M_w = (\frac{\Delta_p}{\Delta S})^2\,\frac{M_s}{2}
\label{eq:sizeweakmeas}
\end{equation} 
The required resources will be supermassive! 

The aspect of weak measurements that has gained great prominence is its alleged \emph{non-invasiveness}. It is clear from 
eqn.(\ref{eq:psiweakpost}) that for low p, the state of the system is practically the same as the unknown intial state. 
In fact, even when $p\cdot s_i \approx \Delta_p^2$, $|\psi(p)\rangle_S$ equals $|\psi\rangle_S$ to a high degree! 
It is instructive to see how the expectation value in eqn.(\ref{eq:weakmean}) gets saturated as the range of outcomes is increased($f\Delta_p
$ is the maximum magnitude of p):
\begin{equation}
R_{sat}=\frac{\langle p \rangle_\psi^f}{\langle p \rangle_\psi} = erf(f) - \frac{2f}{\sqrt{\pi}}\,e^{-f^2}
\label{eq:weakmeansat}
\end{equation}
Here erf is the Gaussian error function. At f=0.5, this ratio is 0.08, at f=1 it is 0.43 while it already reaches 0.94 at f=2! Thus when 
$p\,s_i\approx \Delta_p^2$, values of outcomes where the state remains unaffected to a high degree, the expectation value will be
indistuinguishable from its true value!
These considerations are further strengthened
by looking at the \emph{post-measurement reduced density matrix} of the system:
\begin{equation}
\rho^{post}_S = \rho^{ini} - \frac{1}{4\Delta_p^2}\,\sum_{i,j}\,(s_i-s_j)^2\,\alpha_i\alpha_j^*\,|s_i\rangle \langle s_j|
\label{eq:weakpostreduced}
\end{equation}
giving the impression that the weak measurements are \emph{non-invasive} to a very high degree. The non-invasiveness of weak measurements has
been argued to be useful, for example, in the context of the \emph{Leggett-Garg Inequalities} \cite{leggettgarg1985,homeleggett2013}. One has to ensure the legitimacy of using
\emph{ensembles} in this context and carefully analyse the effect of errors in the light of our remarks earlier. 

The maintenance of the state to such a high degree may give rise to the hope that it holds even for repeated measurements on
single copies. 
One would then have arrived at a way of obtaining \emph{full
information} about a single copy of a system in an \emph{unknown} state without appreciably disturbing it. That would be in conflict with
the \emph{Copenhagen Interpretation}. 
We now show that a careful analysis of the errors in a weak measurement nullify this expectation 
of non-invasiveness. The errors in weak measurements, though highly negligible in a single act of measurement, get so
amplified with repetition as to almost totally disturb the system. This can be heuristically grasped from eqn.(\ref{eq:weakpostreduced})
on recognizing, from eqn.(\ref{eq:sizeweakmeas}), that the number of repetitions must far exceed $\Delta_p^2$ for acceptable error levels;
compounding the change in reduced density matrix per step is seen to totally alter the system state. But eqn.(\ref{eq:weakpostreduced})
is strictly valid only for ensemble measurements. We remedy that in the rest of the paper by working out the consequences of \emph{repeated
weak measurements} on a single copy. 
\section{Repeated weak measurements on a single copy}
Continuous and repeated measurements are well known concepts. For example, they are treated extensively in \cite{jacobs2006,mdsbook2001}.
Sequential weak measurements of several observables are also discussed in \cite{popescuseq2007}. 
The following schema defines for us repeated weak measurements of the same observable on a single copy: (i) perform a weak measurement of system observable S
in state $|\psi\rangle_S$
with the apparatus in the state of eqn.(\ref{eq:iniapp}) with \emph{very large} $\Delta_p$, 
, ii) let the definitive outcome, defined as above, be $p_1$, and the single system state be $|\psi(p_1,\{\alpha\})\rangle_S$, iii) restore 
the apparatus to its initial state, and, 
iv) repeat step (i), and so on. After N such steps, let the sequence of outcomes
be denoted by $p_1,p_2\ldots,p_N$ and the resulting system state by $|\psi(\{p\},\{\alpha\})\rangle_S$.

The probability distribution for the first outcome $p_1$,$P^{(1)}(p_1)$ is simply given by $|N^{(1)}(p_1,\{\alpha\})|^2=|N(p_1,\{\alpha\})|^2$ with $N(p,\{\alpha\})$
given by eqn.(\ref{eq:Nweakpost}). The corresponding system state is given by $|\psi(p_1,\{\alpha\})\rangle_S$ of eqn.(\ref{eq:psiweakpost}). Thus the set of $\alpha$
for this state is given by
\begin{equation}
\alpha_i^{(1)} = \frac{N}{N(p_1,\{\alpha\})}\,e^{-\frac{(p_1-s_i)^2}{2\Delta_p^2}}\,\alpha_i
\label{eq:psiweakfirst}
\end{equation}
Since in step (iii) the apparatus state has been restored, the probability distribution $P^{(2)}(p_2)$ for the outcome $p_2$ at the end of
the second weak measurement, is given by 
\begin{equation}
P^{(2)}(p_2) = 
|N^{(2)}(p_2,\{\alpha\})|^2=|N^{(1)}(p_2,\{\alpha^{(1)}\})|^2 
\end{equation}
Substituting from eqn.(\ref{eq:psiweakfirst}), one gets
\begin{equation}
P^{(2)}(p_2) = \frac{(N^2)^2}{P^{(1)}(p_1)}\,\sum_i\,|\alpha_i|^2\prod\limits_{j=1}^2e^{-\frac{(p_j-s_i)^2}{\Delta_p^2}}
\label{eq:weakprob2}
\end{equation}
It is important to recognize that $P^{(2)}(p_2)$ is actually the \emph{conditional probability} $P(p_2|p_1)$ of obtaining $p_2$
conditional to having already obtained $p_1$ (that is the reason for the explicit dependence on $p_1$ in eqn.(\ref{eq:weakprob2})).
The \emph{joint probability} distribution $P(p_1,p_2)$ is therefore given by $P(p_2,p_1) = P(p_2|p_1)P(p_1)$ to give
\begin{equation}
P(p_1,p_2) = (N^2)^2\,\sum_i|\alpha_i|^2\,\prod\limits_{j=1}^2\,e^{-\frac{(p_j-s_i)^2}{\Delta^2}}
\label{eq:probweak2}
\end{equation}
The state after the second measurement is given by the exact analog of eqn.(\ref{eq:psiweakfirst}):
\begin{equation}
\alpha_i^{(2)} = \frac{N}{N^{(2)}(p_2,\{\alpha^{(1)}\})}\,e^{-\frac{(p_2-s_i)^2}{2\Delta_p^2}}\,\alpha_i^{(1)}
\label{eq:psiweaksec}
\end{equation}
It is useful to explicitly write this state:
\begin{equation}
|\psi(p_1,p_2,\{\alpha\}) = \frac{\sum\limits_i\,\prod\limits_{j=1}^2\,e^{-\frac{(p_j-s_i)^2
}{2\Delta_p^2}}\,\alpha_i|s_i\rangle_S}{\sqrt{\sum\limits_i\,|\alpha_i|^2\prod\limits_{j=1}^2 e^{-\frac{(p_j-s_i)^2}{\Delta_p^2}}}}
\label{eq:psiweak2}
\end{equation}
It is remarkable that these results are all symmetric in the outcomes $p_i$.
Eqns.(\ref{eq:probweak2},\ref{eq:psiweaksec}) readily generalize to the case of M repeated measurements:
\begin{equation}
P(p_1,\ldots,p_M) = (N^2)^M\,\sum_i|\alpha_i|^2\,\prod\limits_{j=1}^M\,e^{-\frac{(p_j-s_i)^2}{\Delta^2}}
\label{eq:probweakN}
\end{equation}
\begin{equation}
|\psi(p_1,\ldots,p_M,\{\alpha\}) = \frac{\sum\limits_i\,\prod\limits_{j=1}^M\,e^{-\frac{(p_j-s_i)^2
}{2\Delta_p^2}}\,\alpha_i|s_i\rangle_S}{\sqrt{\sum\limits_i\,|\alpha_i|^2\prod\limits_{j=1}^M e^{-\frac{(p_j-s_i)^2}{\Delta_p^2}}}}
\label{eq:psiweakN}
\end{equation}
Quite a different approach is taken, for example, by Gurvitz \cite{gurvitz1997}, and by Korotkov \cite{korotkovprb60,korotkovprb63} from the 
formalism used here. 
It is important to understand the precise relationship between these.
The schema used here has been experimentally realized in \cite{katz2006,vijay2012}.
\subsection{Consequences}
The \emph{intrinsic randomness} of quantum theory makes no aspect of a \emph{particular realization}
predictable. For ensemble measurements the variables are \emph{independently}
distributed and the \emph{Central Limit Theorem} guarantees that as long as the number of trials is large enough, averages over even particular
realizations converge nicely to the true mean. To see what happens in the present context, where the outcomes are clearly not 
independently distributed, let us study
y, the average of the M outcomes. The expectation value of y in the joint probability distribution $P(p_1,\ldots,p_M)$
is
\begin{equation}
{\bar y}_M = \frac{1}{M}\,\int\ldots\int\,\prod\limits_{i=1}^M\,\sum_i\,p_i\,P(\{p\}) = \sum_i\,|\alpha_i|^2\,s_i
\label{eq:weakrepeatmean}
\end{equation}
Which is certainly a remarkable result. 
The variance in y can likewise be calculated and it equals $\frac{\Delta_p}
{\sqrt{2M}}$. Thus M has to be chosen according to eqn.(\ref{eq:sizeweakmeas}).

The expectation values and variances are only the \emph{tips of the iceberg} of a distribution. Let us calculate the distribution
function P(y). Though the outcomes are not independently generated, it is nevertheless possible to explicitly calculate this:
\begin{equation}
P(y_M) = \int\ldots\int\,\prod\limits_{i=1}^Mdp_i\,P(\{p\})\delta(y_M-\frac{\sum\limits_i p_i}{M})
\label{eq:weakclt}
\end{equation}
Using eqn.(\ref{eq:probweakN}), this becomes
\begin{equation}
P(y_M) = \sqrt{\frac{M}{\pi\Delta_p^2}}\,\sum\limits_i|\alpha_i|^2\,e^{-\frac{(y_M-s_i)^2M}{\Delta_p^2}}
\rightarrow \sum\limits_i\,|\alpha_i|^2\,\delta(y_M-s_i)
\label{eq:weakclt2}
\end{equation}
where we have also displayed the limiting behaviour as $M\rightarrow\infty$. 

Thus, unlike in the case of ensemble measurements(both strong and
weak), the distribution of $y_M$ is no longer peaked at the true average, with errors decreasing as $M^{-1/2}$. Instead, it is a weighted sum of sharp
distributions peaked around \emph{the eigenvalues}, exactly as in the strong measurement case. 
In other words, averages over outcomes of a particular outcome will be eigenvalues, occurring randomly but with probability $|\alpha_i|^2$.
It then follows that averages over outcomes of a particular realization do not give any information about the initial state!Ensemble
measurements again become inevitable. The other 
consequence is that a very large number of repeated weak measurements on a single copy has the same invasive effect as a strong measurement.
This can also be seen by examining the
expectation value of the system reduced density matrix, $\rho_>^{rep}$: 
\begin{equation}
\rho_>^{rep} = \rho -\sum\limits_{i,j}\,\alpha_i\alpha_j^*\,(1-e^{-\frac{M(s_i-s_j)^2}{4\Delta_p^2}})|s_i\rangle\langle s_j|
\label{eq:weakreprho}
\end{equation}
It is seen
that as M gets larger and larger, there is significant change in the system state. In the limit $M\rightarrow \infty$, the off-diagonal
parts of the density matrix get completely quenched, as in decoherence, and the density matrix takes the diagonal form in the eigenstate of S
basis:
\begin{equation}
\rho_>^{rep}\rightarrow \sum\limits_i\,|\alpha_i|^2|s_i\rangle\langle s_i|
\label{eq:weakrholimit}
\end{equation}
Remarkably, this is exactly the post-measurement density matrix in the case of a strong measurement! 
This decoherence in eigenstate basis of the system has nothing to do with the environmental decoherence in the pointer state basis of the
apparatus. It is a pure manifestation of the repeated measurements. Such an effect was also noted and discussed in
\cite{gurvitz1997}.

It is useful to view these results from the perspectives of error and disturbance. If we take ${\cal D} = 1-tr \rho\cdot\rho_>^{rep}$ as a measure of the disturbance, equivalently the invasiveness, we can
quantify the disturbance in a precise way as a function of the \emph{error} $\epsilon = \frac{\Delta_p}{\sqrt{2M}}$:
\begin{equation}
{\cal D}(\epsilon) = \sum\limits_{i,j}\,|\alpha_i|^2|\alpha_j|^2\,(1-e^{-\frac{(s_i-s_j)^2}{8\epsilon^2}})\rightarrow\sum\limits_i\,|\alpha_i|^2(1-|\alpha_i|^2)
\label{eq:weakerrdist}
\end{equation}
Thus, attempts at reducing errors can only be at the cost of
greater invasiveness. This error-disturbance relation is of a very different nature from those pioneered by Ozawa \cite{ozawa2014}.

The sequence of system states of eqn.(\ref{eq:psiweakN}) is a \emph{random walk} on the state space of the system(see also \cite{korotkovprb60}). It follows from 
eqn.(\ref{eq:psiweakpost}) that the eigenstates of S are the \emph{fixed points} of the probabilistic map that generates this walk.
Presumably each walk terminates in one of the eigenstates but which eigenstate it terminates in is unpredictable. The surprising value
for the mean in eqn.(\ref{eq:weakrepeatmean}) is the result of further 'super-averaging' over a large ensemble of $y_M$. We have only
followed a \emph{frequentist} approach here; it is highly instructive to examine the issues also from a Bayesian perspective.

Alter and Yamomoto have obtained a number of very significant results about the possibility of obtaining information about single
uantum systems \cite{alterbook,alterzeno,alterqnd}. In particular they also gave an analysis based on joint and conditional probabilities
applied to \emph{repeated weak QND} measurements on a single state \cite{alterqnd}. They too obtained evolutions resembling random walks in state
space. They concluded that it is not possible to obtain any information on unknown single states from the statistics of repeated
measurements. The degradation of the state and relation to projective measurements were not explicitly studied. In another work, they found connections between \emph{Quantum Zeno Effect} and the problem of repeated measurements
and again concluded that it is impossible to determine the quantum state of a single system. Our results on information cloning and the
general results from optimal cloning discussed in the next two sections that it may be possible to obtain partial results.

In a very interesting approach to these ontological questions, Paraoanu has investigated these issues within what he calls \emph{partial measurements} \cite{sorinpartial,sorinrepeat}. By employing
a combination of repeated such measurements on a single state and the possibility of reversing such measurements, he too has concluded
the impossibility of obtaining any information about single unknown states. The invasive aspects as well as the connections to strong 
measurements are not explored here either.

{\bf Note added}: This paper first appeared on the quant-ph arxiv(1406.0270) on 19 June 2014. While it was under review by a journal, an
article by Eliahu Cohen, Boaz Tamir and Avner Priel appeared on the same arxiv on 12 Jan 2015(1501.02182) addressing similar issues 
and reaching some of the conclusions reached here especially the convergence of repeated weak measurements to strong measurements.
Their work was subsequently published in \cite{cohenpriel2015}.
But they only treated qubits and their proof of convergence was numerical, whereas our treatment holds for arbitrary quantum systems
with finite dimensional state spaces, and our work is entirely analytical.

\acknowledgments{The author acknowledges warm hospitality at TIFR-TCIS, Hyderabad where this work was done. He thanks Rama Govindarajan for
a perceptive remark on conditional probabilities that simplified the approach to the problem considerably. He thanks M.D. Srinivas for an
incisive discussion on repeated measurements. He acknowledges support from DST, India for the project IR/S2/PU-001/2008.
}

\end{document}